# Transformations of metric tensor and interactive theory of gravity

Shubhen BISWAS

Université de Tours, Parc de Grandmont, 37200, France.

Email: - shubhen3@gmail.com

**Abstract:** In this paper it is reconciled how the metric in Minkowskian space-time gets transformed from one coordinates system to another after successive Lorentz transformations. And likewise this idea is generalized to achieve metric transformation from one curved spacetime to another. The synergy between different sources of masses manifests curved space-time perturbed and the perturbed metric is assumed as point-to-point infinitesimal quasi Lorentz transformation of the Minkowski metric. For two body system this model of linear perturbation interactive gravity is tested for the unstable planetary elliptical orbit and the precession of perihelion is revisited.

**Keywords:** space-time metric, Lorentz boost, linearized metric, modified gravity

**Introduction:** In general relativity and in its application it is extensively studied only for the space-time curvature for single massive body and the dynamics of the test body is presented considering it like a small point mass. If we took the test mass enough massive the scenario would be quite different. As now the synergy between two massive bodies would generate different thing. Here in the following we will proceed to define the shift from usual Newtonian gravity to some extent of modification over linearized metric transformation in curved space-time.

In the four dimensional flat space-time or Minkowskian space-time the invariant line element ($ds^2$) [1, 2]

$$ds^2 = \eta_{\mu\nu} dx^\mu dx^\nu \qquad (1)$$

In matrix form equation (1), we can infer for coordinates systems, $\mathbf{X}$ and $\mathbf{X'}$

$$dX^T[\eta]dX = dX'^T[\eta']dX' \qquad (2)$$

Column vectors, $dX = \begin{pmatrix} dx^0 \\ dx^1 \\ dx^2 \\ dx^3 \end{pmatrix}$; $[\eta] = \begin{pmatrix} -1 & 0 & 0 & 0 \\ 0 & 1 & 0 & 0 \\ 0 & 0 & 1 & 0 \\ 0 & 0 & 0 & 1 \end{pmatrix}$

Also the Lorentz transformations by using Lorentz boost [3, 4] deals with the coordinate transformations

$$dX' = [L_1]dX \qquad (3)$$

The Lorentz transformation matrix, $[L_1] = \begin{pmatrix} \gamma_1 & -\gamma_1\beta_1 & 0 & 0 \\ -\gamma_1\beta_1 & \gamma_1 & 0 & 0 \\ 0 & 0 & 1 & 0 \\ 0 & 0 & 0 & 1 \end{pmatrix}$; $\gamma_1 = \frac{1}{\sqrt{1-\frac{v_1^2}{c^2}}}$; $\beta_1 = \frac{v_1}{c}$;

$v_1$ uniform velocity unidirectional axis of $X'$ relative to $X$.

The metric transformations under Lorentz boost by plugging equation (3) to equation (2)

$$dX^T[\eta^{(1)}]dX = dX^T[L_1]^T[\eta'][L_1]dX \qquad (4)$$

We have the transformed metric for any isolated transformations

$$[\eta^{(i)}] = [L_i]^T[\eta'][L_i] \qquad (5)$$

Equation (5) is the Lorentz invariance of metric tensor in one space-time to another. In case of two consecutive transformations, $\rightarrow \mathbf{X}' \rightarrow \mathbf{X}''$, invariance still holds good.

$$[\eta] = [L_2]^T [L_1]^T [\eta'][L_1]\,[L_2] \qquad (6)$$

This metric transformation is done for coordinate transformations in velocity space and is a linear transformation in Euclidean space.

In a Non-Euclidean style **[3, 4]** if we go for transformations in rapidity space **[3]** the equation (3) becomes

$$[dX] = \mathbf{\Lambda}(w)[dX'] \qquad (7)$$

$$\mathbf{\Lambda}(w) = e^{Zw}\ ;\ \mathbf{Z} = \begin{bmatrix} 0 & -1 \\ -1 & 0 \end{bmatrix}\ ;\ w_1 = -\ln[\gamma_1(1-\beta_1)]\ ;\ [\Lambda] = \left[\mathbb{1} + Zw + \frac{(Zw)^2}{2!} + \cdots\right]$$

Equation (7) turns transformations into nonlinear fashion, indulges for achieving suitable transformations in the Non-Euclidean curved space-time.

## 2. Generalization of flat space-time metric transformations to curved space-time metric:

From Einstein's principle of equivalence the effects of gravity is equivalent to the experience in a curved space-time **[5, 6, 7]**. Introduction of massive body in a flat space-time or Minkowskian space-time **[1]** turns out it to curved space-time. The weak-field approximation presumes "nearly Cartesian" coordinates, and line elements is given by

$$ds^2 = g_{\mu\nu}dx^\mu dx^\nu \qquad (8)$$

Where the metric, $[g_{\mu\nu}] = [\eta_{\mu\nu} + h_{\mu\nu} + \text{ (higher order terms in } h_{\mu\nu})]$ \qquad (9)

Choosing orthogonal coordinate system, the flat space-time Minkowski metric **[8]** and with the linear perturbation in matrix form as the standard post-Newtonian form (PPN) in local quasi-Cartesian coordinates **[2]**,

$$[h_{\mu\nu}] = \begin{pmatrix} h_{00} & 0 & 0 & 0 \\ 0 & h_{11} & 0 & 0 \\ 0 & 0 & h_{22} & 0 \\ 0 & 0 & 0 & h_{33} \end{pmatrix}$$

From principle of equivalence we have the freedom to choose "locally inertial coordinates system" **[1, 7]** at every space-time point in an arbitrary gravitational field. Owing to this in weak gravitational field we can think for curved space time metric as the point to point infinitesimal quasi Lorentz transformations of Minkowski metric. This can be done in a little tricky way, where initial metric that is the Minkowski metric is placed as the multiplicative factor of metric deformation. Such that the local coordinates system faced quasi-Lorentz transformations with a factor of initial metric

$$[g_{\mu\nu}{}^{(1)}] = \left[\mathbb{1} + [h_{\mu\nu}{}^{(1)}][\eta_{\mu\nu}{}^{(1)}]^{-1}\right][\eta_{\mu\nu}{}^{(1)}] \qquad (10)$$

Using equation (5) as the transformation in local space time metric

$$[g_{\mu\nu}{}^{(1)}] = \mathbf{\Lambda}_1 [L_1]^T [\eta'_{\mu\nu}]\,[L_1] \qquad (11)$$

$$\mathbf{\Lambda}_1 = \left[\mathbb{1} + [h_{\mu\nu}{}^{(1)}][\eta_{\mu\nu}{}^{(1)}]^{-1}\right]\ ;\ [L_1]\text{symmetric denoting, } \boldsymbol{\mathcal{L}}_1 = \mathbf{\Lambda}_1[L_1]^T = \mathbf{\Lambda}_1 \boldsymbol{L}_1$$

$$[g_{\mu\nu}{}^{(1)}] = \boldsymbol{\mathcal{L}}_1 [\eta'_{\mu\nu}]\,\boldsymbol{L}_1 \qquad (12)$$

The equation (12) stands for single quasi Lorentz transformation. Where neglecting higher order perturbation term, the space-time metric in Post-Newtonian Non-Euclidean curved space-time for weak gravity theory is nearly Cartesian, that is perturbed over flat space-time [5, 7, 8, 9],

$$[h_{\mu\nu}][\eta_{\mu\nu}]^{-1} = \begin{pmatrix} -h_{00} & 0 & 0 & 0 \\ 0 & h_{11} & 0 & 0 \\ 0 & 0 & h_{22} & 0 \\ 0 & 0 & 0 & h_{33} \end{pmatrix} \qquad (13)$$

For the binary system the fabric of background spacetime for the mass $'M'$ (source1) is perturbed by the weak field self gravity of the mass $'m'$ (source 2). We can here consider the initial metric get changes as the same fashion as in equation (6), unlike following successive two quasi-Lorentz transformations implying equation (12) in non-Euclidean curved space-time.

$$[g_{\mu\nu}] = \mathcal{L}_2 \mathcal{L}_1 \left[\eta'_{\mu\nu}\right] L_1 L_2 \qquad (14)$$

$$[g_{\mu\nu}] = \Lambda_2 L_2 \Lambda_1 L_2^{-1} \left[\eta_{\mu\nu}\right] \qquad (15)$$

For transformations in weak field approximation the local inertial coordinates system allows,

$$h_{\mu\nu}\beta_2 = h_{\mu\nu}\frac{v_2}{c} \sim 0 \text{ and } \gamma_2 \sim 1, \text{ we have } L_2 \Lambda_1 L_2^{-1} = \Lambda_1$$

$$[g_{\mu\nu}] = \Lambda_1 \Lambda_2 [\eta_{\mu\nu}] \qquad (16)$$

$\Lambda_1$ and $\Lambda_2$ are diagonal matrices

$$\Lambda_2 \Lambda_1 = \Lambda_1 \Lambda_2 = \left[\mathbb{1} + \left[h_{\mu\nu}^{(1)}\right]\left[\eta^{(1)}{}_{\mu\nu}\right]^{-1}\right]\left[\mathbb{1} + \left[h_{\mu\nu}^{(2)}\right]\left[\eta^{(2)}{}_{\mu\nu}\right]^{-1}\right] \qquad (17)$$

The equation (16) gives the new perturbed metric for two different massive sources as multiplication of one isolated background metric with the others metric deformation. Also equation (16) is obvious in regard that offing any one of the sources from the system reproduces the corresponding metric for the existing single source in the linearized perturbation theory.

For $n$ sources using the law of induction transformed metric is

$$[g_{\mu\nu}] = \prod_{i=1}^{n}\left[\mathbb{1} + \left[h_{\mu\nu}^{(i)}\right]\left[\eta^{(i)}{}_{\mu\nu}\right]^{-1}\right][\eta_{\mu\nu}] \qquad (18)$$

## 3. Effective potential for two body system:

The static isotropic metric solution of the equations is given by Schwarzschild [5, 7, 9, 10, 11, 12, 13 and14]

$$ds^2 = -\left[1 + \frac{2\varphi}{c^2}\right]c^2 dt^2 + \left[1 + \frac{2\varphi}{c^2}\right]^{-1} dr^2 + r^2 d\theta + r^2 sin^2\theta d\emptyset^2 \qquad (19)$$

The perturbed component in terms of Newtonian potential from equation (28)

$$h_{00} = -\frac{2\varphi}{c^2} \qquad (20)$$

For a suitable example let us imagine two different masses $'M'$ and '$m$', at positions $r_1$ and r2, from the observation point, and then from equation (18) the perturbed metric has the component

$$g_{00} = -\left[1 - \left(h_{00}^{(1)} + h_{00}^{(2)}\right) + h_{00}^{(1)} h_{00}^{(2)}\right] \qquad (21)$$

$$\Phi = \left(\varphi_1 + \varphi_2 + \frac{2\varphi_1 \cdot \varphi_2}{c^2}\right) \qquad (22)$$

Gravitational potentials $\varphi_1 = -\frac{GM}{r_1}$ and $\varphi_2 = -\frac{Gm}{r_2}$

$$\Phi = -\left(\frac{GM}{r_1} + \frac{Gm}{r_2}\right) + \frac{2G^2Mm}{c^2 r_1 r_2} \qquad (23)$$

The formula (23) clearly indicates a shift, an additive interaction term with Newton potential. This additional term due to linearized perturbation interactive gravity can describe the quasi stable orbit **[7, 15, 16]** and the precession of perihelion **[8, 10, 16, 17, 20]** in detail as ***Appendix-A***.

## 4. Conclusions:

In the light of linearized perturbation for multiple sources the metric tensor is given by the equation (18). This formula is tested for two body case where it keeps an additional potential term with the pure inverse Newton gravitational potentials, shows a significant shift in Newtonian gravity for two massive bodies. The formula (23) has an additional interactive term besides the additive nature of potentials. Also single massive source doesn't result any extra term following pure Newtonian gravity. This quite phenomenal anomaly comes from the synergy between two massive sources. This model of modification of Newtonian gravity reproduces the expected outcome as the case of the perihelion precession for the two body system in the quasi stable planetary orbits (***Appendix-A***). The precession energy is sourced from the non Newtonian additive two body interaction.

**Acknowledgement:** I am especially thankful to Prof. Stam NICOLIS encouraged me throughout in writing the paper.

**Data Availability Statement:**

Data sharing not applicable to this article as no datasets were generated or analysed during the current study.

**Declarations:**

For this work there is no Funding and/or Conflicts of interests/Competing interests.

### *Appendix-A*

**The planetary perihelion precession:** Rewriting equation (23) in the Newton form the gravitational potential $\Phi$ at the observation point 'O' as in **Fig.1**

$$\Phi = -\frac{GM_G(r_2)}{r_1} - \frac{Gm_G(r_1)}{r_2} \qquad (24)$$

From their perspective to the inertial masses $'M'$ and $'m'$ the equation (24) is for attributing gravitational masses $M_G(r)$ and $m_G(r)$ respectively, in presence of their mutual interaction.

$$M_G(r) = M\left(1 - \frac{Gm}{c^2 r}\right) \qquad (25)$$

$$m_G(r) = m\left(1 - \frac{GM}{c^2 r}\right) \qquad (26)$$

The form (25) and (26) is not ignorable for massive bodies and describes a shift to the form of the inverse Newtonian gravitational potential energy $V(r)$.

In the case of time like geodesic using the Schwarzschild metric the normalized Lagrangian($\epsilon = -1$); $\mathfrak{L} = g_{\mu\nu}\dot{x}^\mu\dot{x}^\nu = \epsilon$, gives the energy equation **[8, 17]** ( $G = c = 1$),

$$\frac{E_s^2 + \epsilon}{2} = \frac{1}{2}\left(\frac{dr}{d\tau}\right)^2 + \left(\epsilon\frac{M}{r} + \frac{l^2}{2r^2} - \frac{Ml^2}{r^3}\right) \qquad (27)$$

Putting back, $\frac{M}{r} \rightarrow \frac{GM}{c^2 r}$ and $\epsilon \rightarrow -mc^2$, considering the mass $'m'$ is sitting fix (taking angular momentum, $l = 0$) at a certain distance $'r'$ from the mass $'M'$ then it shows,

$$E_S/c^2 = m\left(1 - \frac{2GM}{c^2 r}\right)^{\frac{1}{2}} \approx m_G(r) \qquad (28)$$

Thus equation (28) guarantees the consistency of the derived forms in (25) and (26) in accord the general relativistic views and deals with the modification of potential energy associated with mass 'm' and 'M' as the following

$$V^*(r) = -\frac{GMm}{r}\left(1 - \frac{GM}{c^2 r}\right)\left(1 - \frac{Gm}{c^2 r}\right) \qquad (29)$$

In usual case of planetary motion $M \gg m$ , $M + m \approx M$

$$V^*(r) = -\frac{GMm}{r} + \frac{G^2 M^2 m}{c^2 r^2} + \mathcal{O}\left(\frac{G^3 M^2 m^2}{c^4 r^3}\right) \qquad (30)$$

$$\phi^*(r) = -\frac{GM}{r} + \frac{G^2M^2}{c^2r^2} \qquad (31)$$

Finally the modified formula (31) describes a clear departure from pure inverse Newton gravitational potential.

The Lagrangian for central force in Newtonian gravity taking azimuthal angle $\theta$,

$$\mathfrak{L} = \frac{1}{2}\dot{r}^2 + \frac{1}{2}r^2\dot{\theta}^2 + \frac{GM}{r} \qquad (32)$$

In motion equation (32) implies conserved angular momentum

$$r^2\dot{\theta} = l (const. in motion) \qquad (33)$$

Energy equation using polar coordinate two body system in pure Newtonian gravity **[7, 8, 17]**,

$$\frac{1}{2}\dot{r}^2 = \frac{E_N}{m} - \left(\frac{l^2}{2r^2} - \frac{GM}{r}\right) = \frac{E_N}{m} - \frac{U}{m} \quad (34)$$

Effective potential in Newton,

$$\frac{U}{m} = \frac{l^2}{2r^2} - \frac{GM}{r} = \frac{l^2\phi^2}{2G^2M^2} + \phi \qquad (35)$$

For stable orbit, $\frac{d\left(\frac{U}{m}\right)}{dr}\Big|_{r=r_0} = 0$,

$$l^2 = GMr_0 \qquad (36)$$

$$\dot{\theta}^2 = \omega_{\theta_N}^2 = \frac{GM}{r_0^3} \qquad (37)$$

Also for radial oscillation angular velocity in Newton,

$$\frac{d^2\left(\frac{U}{m}\right)}{dr^2}\Big|_{r=r_0} = \omega_{r_N}^2 = \frac{GM}{r_0^3} \qquad (38)$$

So it is expected in changed scenario of effective potential we will get different formulas for all the elements (32) to (38) as in the followings.

From Bertrand's theorem **[7, 15]** inverse Newtonian potential energy($r^{-1}$) deals with closed circular orbit and any little departure from the inverse potential energy led to open orbit.

The formula (31) describe a shift from the inverse Newtonian gravitational potential $\phi(r)$, and we expect deviation from the closed circular stable orbit to nearly circular open unstable orbit **[16]** and its radial time period.

Since the interaction contribute an additional positive energy to the system from its Newtonian counterpart and we can expect the little amount of additional increase in energy will contribute to its total dynamic energy. We need to modify equations (34) and (38) in accordance with deviated Newtonian potential $\phi^*$.

The departure is small enough such that, $l^{*2} = \left[GMr_0 - \frac{2G^2M^2}{c^2}\right] \approx GMr_0$, i.e. angular momentum is almost constant and $\dot{\theta}^2 \approx \frac{GM}{r_0^3}$, unchanged in deviated potential. But it is not ignorable for the case of the Effective potential due to extra potential in the formula (34). Now the equation of energy for radial motion turns in to

$$\frac{1}{2}\left(\frac{dr}{dt}\right)^2 = \frac{E_N^*}{m} - \left(\frac{l^{*2}\phi^{*2}}{2G^2M^2} + \phi^*\right) = \frac{E_N^*}{m} - \frac{U^*}{m} \quad (39)$$

$$\frac{U^*(r)}{m} = \frac{GMr_0}{2r^2} - \frac{G^2M^2r_0}{c^2r^3} + \mathcal{O}\left(\frac{G^3M^3r_0}{2c^4r^4}\right) - \frac{GM}{r} + \frac{G^2M^2}{c^2r^2} \quad (40)$$

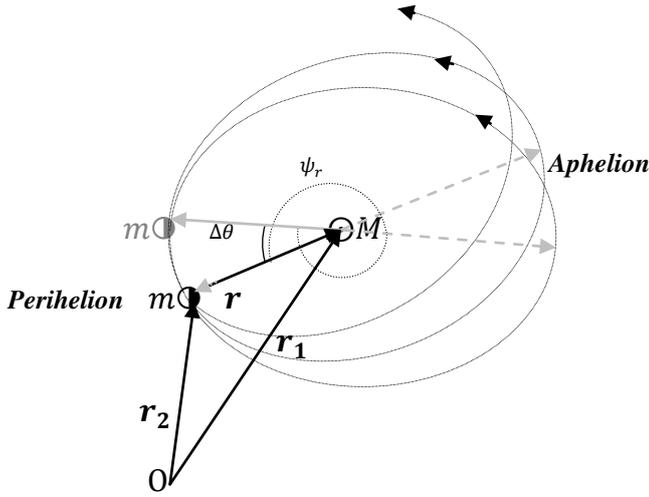

**Fig:-1**
**The perihelion precession of the two body system in unstable orbits**

The equation (40) is enough in computing the radial angular velocity using formula (38). Extra effective potential term is responsible for angular advances of perihelion precession [7, 16, 17].

$$\frac{d^2(U^*/m)}{dr^2}\Big|_{r=r_0} = \omega_r^2 = \frac{GM}{r_0^3}\left[1 - \frac{6GM}{c^2r_0}\right] \qquad (41)$$

$$\omega_r = \omega_\theta\left(1 - \frac{3GM}{c^2r_0}\right) \qquad (42)$$

$\frac{d^2\left(\frac{U^*}{m}\right)}{dr^2}\Big|_{r=r_0} > 0$, this means the orbit is stable [7, 18, 19] or may be quasi stable, no *stable circular orbits exist for* $r \le \frac{6GM}{c^2}$. Under perturbed infinitesimally small it becomes a quasi-circular orbit [19] and $'r'$ will oscillate about $r_0$ with radial time period $T_r$.

$$T_r = T_\theta\left(1 - \frac{3GM}{c^2r_0}\right)^{-1} \qquad (43)$$

For radial time period the angle advances per revolution as,

$$\psi_r \approx \omega_\theta T_\theta\left(1 + \frac{3GM}{c^2r_0}\right) = \left(2\pi + \frac{6\pi GM}{c^2r_0}\right) \qquad (44)$$

In switching over circular to elliptical Keplarian orbit, $r_0 \sim semilatus\ rectum = a(1 - e^2)$, $a$ is the semi major axis and $e$ is the eccentricity of the ellipse. The equation (44) gives precession of perihelion,

$$\psi_r - 2\pi = \Delta\theta \approx \frac{6\pi GM}{c^2 a(1-e^2)} \ radians/period \quad (45)$$